%% file: main.tex
\documentclass[letterpaper, 11pt]{article}

\usepackage{appendix}

\newcommand{\comment}[1]{}

\usepackage{graphicx}
\usepackage{latexsym}
\usepackage{mdwlist}

\usepackage{algorithm}

\newcommand{\BSB}{{\em Broadcast\_Single\_Bit~}}
\newcommand{\DG}{{\em Diag\_Graph~}}
\newcommand{\TRUE}{{\bf true~}}
\newcommand{\FALSE}{{\bf false~}}

\newenvironment{proof}{\paragraph{\bf Proof:}}{\hspace*{\fill}\(\Box\)}
\newtheorem{theorem}{Theorem}

\newtheorem{lemma}{Lemma}

\def\noflash#1{\setbox0=\hbox{#1}\hbox to 1\wd0{\hfill}}

\begin{document}

\title{Error-Free Multi-Valued Consensus with Byzantine Failures
 \footnote{\normalsize This research is supported
in part by Army Research Office grant W-911-NF-0710287 and National
Science Foundation award 1059540. Any opinions, findings, and conclusions or recommendations expressed here are those of the authors and do not
necessarily reflect the views of the funding agencies or the U.S. government.}}

\author{Guanfeng Liang and Nitin Vaidya\\ \normalsize Department of Electrical and Computer Engineering, and\\ \normalsize Coordinated Science Laboratory\\ \normalsize University of Illinois at Urbana-Champaign\\ \normalsize gliang2@illinois.edu, nhv@illinois.edu}

\maketitle
\date{}
\input{abstract}


\thispagestyle{empty}

\newpage

\setcounter{page}{1}

\input{intro}

\input{nutshell}


\input{consensus}
\input{broadcast_short}
\input{conclusion}

\newpage
\bibliography{PaperList}


\end{document}

%% file: abstract.tex
\begin{abstract}
In this paper, we present an efficient {\em deterministic} algorithm for consensus in presence of Byzantine failures. Our algorithm achieves consensus on an $L$-bit value with communication complexity $O(nL + n^4 L^{0.5}  + n^6)$ bits, in a network consisting of $n$ processors with up to $t$ Byzantine failures, such that $t<n/3$. For large enough $L$, communication complexity of the proposed algorithm approaches $O(nL)$ bits. In other words, for large $L$, the communication complexity is linear in the number of processors in the network. This is an improvement over the work of Fitzi and Hirt (from PODC 2006), who proposed a probabilistically correct multi-valued Byzantine consensus algorithm with a similar complexity for large $L$. In contrast to the algorithm by Fitzi and Hirt, our algorithm is guaranteed to be always error-free. Our algorithm require no cryptographic technique, such as authentication, nor any secret sharing mechanism. To the best of our knowledge, we are the first to show that, for large $L$, error-free multi-valued Byzantine consensus on an $L$-bit value is achievable with $O(nL)$ bits of communication.


\end{abstract}

~

~

%% file: intro.tex
\section{Introduction}
\label{sec:intro}
This paper considers the multi-valued Byzantine {\em consensus} problem.
The Byzantine consensus problem considers $n$ processors, namely $P_1,...,P_n$, of which at most $t$ processors may be {\em faulty} and deviate from the algorithm in arbitrary fashion. Each processor $P_i$ is given an $L$-bit input value $v_i$, and they want
to agree on a value $v$ such that the following properties are satisfied:
\begin{itemize}
\item {\em Termination}: every fault-free $P_i$ eventually decides on an output value $v_i'$,
\item {\em Consistency}: the output values of all fault-free processors are equal, i.e.,
for every fault-free processor $P_i$, $v_i'=v'$ for some $v'$,
\item {\em Validity}: if every fault-free $P_i$ holds the same input $v_i=v$ for some $v$, then $v'=v$.
\end{itemize}

\noindent
Algorithms that satisfy the above properties in all executions 
are said to be {\bf error-free}.

We are interested in the communication complexity of error-free consensus algorithms.
 {\em Communication complexity} of an algorithm is defined as the maximum (over all permissible
executions) of the total number of bits transmitted by all the
processors according to the specification of the algorithm.
This measure of complexity was first introduced by Yao \cite{Yao_comm_complexity}, and has been widely used by the distributed computing community  \cite {authenticated_BA_Dolev83,multi-valued_BA_PODC06, Waidner96information-theoreticpseudosignatures}. 

\paragraph{System Model:}
We assume network and adversary models commonly
 used in other related work \cite{psl82, bit_optimal_89, opt_bit_Welch92, multi-valued_BA_PODC06,King:PODC2010}.

We assume a synchronous fully connected network of $n$ processors, wherein the processor identifiers are common knowledge. Every pair of processors are connected with a pair of directed point-to-point communication channels.
Whenever a processor receives a message on such a directed channel, it can correctly assume that the message
is sent by the processor at the other end of the channel.

We assume a Byzantine adversary that has complete knowledge of the state of the other processors, including
the $L$-bit input values.
No secret is hidden from the adversary. 
The adversary can take over up to $t$ processors ($t<n/3$) at any point during the algorithm.
These processors are said to be {\em faulty}. The faulty processors can engage in any 
``{\em misbehavior}'', i.e., deviations from the algorithm, including sending incorrect messages, and
collusion. The remaining processors are {\em fault-free} and follow the algorithm.

Finally, we make no assumption of any cryptographic technique, such as authentication and secret sharing.

%


~

It has been shown that error-free consensus is impossible if $t\ge n/3$ \cite{psl80,psl82}.
$\Omega(n^2)$ has been shown to be a lower bound on the number of messages needed to achieve error-free consensus \cite{Bounds_BA_Dolev85}. Since any message must be of at least 1 bit, this gives a lower bound of $\Omega(n^2)$ bits on the communication complexity of any binary (1-bit) consensus algorithm.



In practice, agreement is sometimes required for longer messages rather than just single bits.
For instance, the ``value'' being agreed upon may be a large file in a
fault-tolerant distributed storage system.
For instance, as \cite{multi-valued_BA_PODC06} suggests, in a voting protocol, the authorities
must agree on the set of all ballots to be tallied (which can be gigabytes of data).
Similarly, as also suggested in \cite{multi-valued_BA_PODC06}, multi-valued Byzantine agreement
is relevant in secure multi-party computation, where many broadcast invocations can be
parallelized and thereby optimized to a single invocation with a long message.

The problem of achieving consensus on a single $L$-bit value may be solved
using $L$ instances of a 1-bit consensus algorithm. However, this approach will
result in communication complexity of $\Omega(n^2L)$, since $\Omega(n^2)$ is a
lower bound on communication complexity of 1-bit consensus.
In a PODC 2006 paper, Fitzi and Hirt \cite{multi-valued_BA_PODC06} presented a probabilistically correct multi-valued consensus algorithm which improves the communication complexity to $O(nL)$ for sufficiently large $L$, at the cost of allowing a non-zero probability of error. Since $\Omega(nL)$ is a lower bound on the communication complexity of consensus on an $L$-bit value, this algortihm has optimal complexity for large $L$.
In their algorithm, an $L$-bit 
value (or message) is first reduced to a much shorter message, using a universal hash function. Byzantine consensus is then performed for the shorter hashed values. Given the result of consensus on the hashed values, consensus on $L$ bits is then achieved by requiring processors whose $L$-bit input value matches the agreed hashed value deliver the $L$ bits to the other processors jointly. By performing initial consensus only for the smaller hashed values, this algorithm is able to reduce the communication complexity to $O(nL+n^3(n+\kappa))$ where $\kappa$ is a parameter of the
algorithm.  However, since the hash function is not collision-free, this algorithm is {\bf not} error-free.
Its probability of error is lower bounded by the collision  probability of the hash function.

We improve on the work of Fitzi and Hirt \cite{multi-valued_BA_PODC06}, and present a deterministic error-free consensus algorithm
with communication complexity of $O(nL)$ bits for sufficiently large $L$. Our algorithm {\em always} produce the correct result, unlike \cite{multi-valued_BA_PODC06}.
For smaller $L$, the communication complexity of our algorithms is  $O(nL + n^4 L^{0.5}  + n^6)$.
To our knowledge, this is the first known error-free multi-valued Byzantine consensus algorithm that achieves, for large $L$, communication complexity linear in $n$.

%% file: nutshell.tex
\section{Byzantine Consensus: Salient Features of the Algorithm}
\label{sec:nutshell}

The goal of our consensus algorithm is to achieve consensus on an $L$-bit value (or message). 
The algorithm is designed to perform efficiently for large $L$. Consequently, our discussion
will assume that $L$ is ``sufficiently large'' (how large is ``sufficiently large'' will become
clearer later in the paper).
We now briefly describe the salient features of our consensus algorithm,
with the detailed algorithm presented later in Section \ref{sec:consensus}.

\begin{itemize}
\item {\em Algorithm execution in multiple generations:}
To improve the communication complexity, consensus on the $L$-bit value is performed
``in parts''. In particular, for a certain integer $D$, the $L$-bit value is divided
into $L/D$ parts, each consisting of $D$ bits.
For convenience of presentation, we will assume that $L/D$ is an integer.
A sub-algorithm is used to perform
consensus on each of these $D$-bit values, and we will refer to each execution of
the sub-algorithm as a ``generation''.

\item {\em Memory across generations:}
If during any one generation, misbehavior by some faulty processor is detected, then
additional (and expensive) diagnostic steps are performed to gain information on the potential identity
of the misbehaving processor(s). This information is captured by means of a {\em diagnosis
graph}, as elaborated later. As the sub-algorithm is performed for each new generation,
the {\em diagnosis graph} is updated to incorporate any new information that may be learnt
regarding the location of the faulty processors. The execution of the sub-algorithm in each
generation is adapted to the state of the diagnosis graph at the start of the generation.

\item {\em Bounded instances of misbehavior:} With Byzantine failures,
it is not always possible
to immediately determine the identity of a misbehaving processor.
However, due to the manner in which the diagnosis graph
is maintained, and the manner in which the sub-algorithm adapts to the
diagnosis graph, the $t$ (or fewer) faulty processors can collectively misbehave in at most $t(t+1)$
generations, before all the faulty processors are exactly identified. Once a faulty processor
is identified, it is effectively isolated from the network, and cannot tamper
with future generations. Thus, $t(t+1)$ is also an upper bound on the number of
generations in which the expensive diagnostic steps referred above may need to be performed.

\item {\em Low-cost failure-free execution:}
Due to the bounded number of generations in which the faulty processors can misbehave, it turns
out that the faulty processors do not tamper with the execution in a majority of the generations.
We use a low-cost mechanism to achieve consensus in failure-free generations, which helps
to achieve low communication complexity. In particular, we use an {\em error detecting code}-based strategy
to reduce the amount of information the processors must exchange to be able to achieve
consensus in the absence of any misbehavior (the strategy, in fact, also allows detection
of potential misbehavior).

\item {\em Consistent diagnosis graph maintenance:}
A copy of the diagnosis graph is maintained locally by each fault-free processor.
To ensure consistent maintenance of this graph, the {\em diagnostic information}
(elaborated later) needs to be distributed consistently to all the processors
in the network. This operation itself requires a Byzantine broadcast algorithm that solves the ``Byzantine Generals Problem'' \cite{psl82}. With this algorithm, a ``source'' processor broadcasts its message to all other processors reliably, even if some processors (including the source) may be faulty.
For this operation we use an error-free 1-bit
Byzantine broadcast algorithm that tolerates $t<n/3$ Byzantine failures with communication complexity of $O(n^2)$ bits  \cite{opt_bit_Welch92,bit_optimal_89}.
This 1-bit broadcast algorithm is referred as \BSB in our discussion. While \BSB is expensive,
the cumulative overhead of \BSB is kept low by invoking it a relatively small number of times, when
compared to $L$.
\end{itemize}

We now elaborate on the error detecting code used in our algorithms, and also
describe the {\em diagnosis graph} in some more detail.

\paragraph{Error detecting code:}
We will use Reed-Solomon codes in our algorithms (potentially other
codes may be used instead). Consider a $(m,k)$ Reed-Solomon code
in Galois Field GF($2^c$), where $c$ is chosen large enough (specifically, $m \leq 2^c - 1$).
This code encodes $k$ data symbols from GF($2^c$) into a codeword consisting
of $m$ symbols from GF($2^c$).
Each symbol from GF($2^c$) can be represented using $c$ bits. Thus,
a data vector of $k$ symbols contains $kc$ bits, and the corresponding codeword
contains $mc$ bits.

Each symbol of the codeword is computed as a linear combination of the $k$ data symbols,
such that every subset of $k$ coded symbols represent a set of linearly independent combinations
of the $k$ data symbols. This property implies that any subset of $k$ symbols from the $m$ symbols of
a given codeword can be used to determine the data vector corresponding to the codeword.
Similarly, knowledge of any subset of $k$ symbols from a codeword suffices to determine the
remaining symbols of the codeword. So $k$ is also called the {\em dimension} of the code.

For a code $C$, let us denote $C()$ as the encoding function, and $C^{-1}()$ as the
decoding function. The decoding function can be applied so long as at least $k$
symbols of a codeword are available. 


\paragraph{Diagnosis Graph:}
The fault-free processors' (potentially partial) knowledge of the identity of the
faulty processors is captured by a diagnosis graph. A diagnosis graph is an undirected graph
with $n$ vertices, with vertex $i$ corresponding to processor $P_i$. A pair of processors are
said to ``trust'' each other if the corresponding pairs of vertices in the diagnosis
graph is connected with an edge; otherwise they are said to ``accuse'' each other. 

Before the start of the very first generation,
the diagnosis graph is initialized as a fully connected graph, which implies that all
the $n$ processors initially trust each other.
During the execution of the algorithm, whenever misbehavior by some faulty processor
is detected, the diagnosis graph will be updated, and one or more edges  will be removed
from the graph, using the diagnostic information communicated using the \BSB algorithm.
The use of \BSB ensures that the fault-free processors always have a consistent view of
the diagnosis graph. As we will show later, the evolution of the diagnosis graph satisfies
the following properties:
\begin{itemize}
\item If an edge is removed from the diagnosis graph, at least one of the processors corresponding to
the two endpoints of the removed edge must be faulty.
\item The fault-free processors always trust each other throughout the algorithm.
\item If more than $t$ edges at a vertex in the diagnosis graph are removed, then the processor corresponding to that vertex must
be faulty.
\end{itemize}
The last two properties above follow directly from the first property,
and the assumption that there are at most $t$ faulty processors.


%% file: consensus.tex
\section{Multi-Valued Consensus}
\label{sec:consensus}

In this section, we describe our consensus algorithm, present a proof of correctness.

The $L$-bit input value $v_i$ at each processor is divided into $L/D$ parts 
of size $D$ bits each, as noted earlier. These parts are denoted as
$v_i(1), v_i(2), \cdots , v_i(L/D)$.

Our algorithm for achieving $L$-bit consensus
consists of $L/D$ sequential executions of Algorithm \ref{alg:consensus}
presented in this section (we will discuss the algorithm in detail below). Algorithm \ref{alg:consensus}
is executed once for each generation. For the $g$-th generation ($1\le g \le L/D$),
each processor $P_i$ uses $v_i(g)$ as its input in Algorithm \ref{alg:consensus}.
Each generation of the algorithm results in processor $P_i$ deciding on
$g$-th part (namely, $v_i'(g)$) of its final decision value $v_i'$.

The value $v_i(g)$ is represented by a vector of $n-2t$ symbols,
each symbol represented with $D/(n-2t)$ bits. For convenience of presentation,
we assume that $D/(n-2t)$ is an integer.
We will refer to these $n-2t$ symbols as the {\em data symbols}.

A $(n,n-2t)$ distance-$(2t+1)$ Reed-Solomon code, denoted as $C_{2t}$, is used to encode the
$n-2t$ data symbols into $n$ coded symbols.
We assume that $D/(n-2t)$ is large enough to allow the above Reed-Solomon code to exist,
specifically, $n \le 2^{D/(n-2t)} - 1$. This condition is met only if $L$ is large enough
(since $L > D$).

We now present some notations to be used in our discussion below.
For a $m$-element vector $V$, we denote $V[j]$ as the $j$-th element of the vector, $1\le j \le m$.
Given a subset $A \subseteq \{1,\dots,m\}$, denote $V/A$ as the ordered
list of elements of $V$ at the locations corresponding to elements of $A$.
For instance, if $m=5$ and $A=\{2,4\}$, then $V/A$ is equal to $(V[2],V[4])$.
We will say that $V/A \in C_{2t}$ if
there exists a codeword $Z\in C_{2t}$ such that $Z/A = V/A$. Otherwise, we will say
that $V/A \notin C_{2t}$.
Suppose that $Z$ is the codeword corresponding to data $v$. This is denoted
as $Z = C_{2t}(v)$, and $v = C_{2t}^{-1}(Z)$.
We will extend the definition of the inverse function $C_{2t}^{-1}$ as follows.
When set $A$ contains at least $n-2t$ elements, we will define 
$C_{2t}^{-1}(V/A) = v$, if there exists a codeword $Z\in C_{2t}$ such that
$Z/A = V/A$ and $C_{2t}(v) = Z$.

Let the set of all the fault-free processors be denoted as $P_{good}$.

Algorithm \ref{alg:consensus} for each generation $g$ consists of three stages.
We summarize the function of these three
stages first, followed by a more detailed discussion:
\begin{enumerate}
\item Matching stage: Each processor $P_i$ encodes its $D$-bit input $v_i(g)$ for generation $g$ into $n$ coded symbols,
as noted above. Each processor $P_i$ sends one of these $n$ coded symbols to the other processors
{\bf that it trusts}. Processor $P_i$ trusts processor $P_j$ if and only if the corresponding vertices in the
diagnosis graph are connected by an edge.
Using the symbols thus received from each other, the processors attempt to identify a ``matching set'' of processors
(denoted $P_{match}$) of size $n-t$ such that the fault-free processors in $P_{match}$ are guaranteed to have
an identical input value for the current generation. If such a $P_{match}$ is not found, it can be
determined with certainty that all the fault-free processors do not have the same input value -- in this case,
the fault-free processors decide on a default output value and terminate the algorithm.

\item Checking stage: If a set of processors $P_{match}$ is identified in the above
matching stage, each processor $P_j \notin P_{match}$ checks whether
the symbols received the from processors in $P_{match}$ correspond to a valid codeword.
If such a codeword exists, then the symbols received from $P_{match}$ are said
to be ``consistent''.
If any processor finds that these symbols are not consistent, then
misbehavior by some faulty processor is detected. 
Else all the processors are able to correctly compute the value to be agreed upon in the
current generation.

\item Diagnosis stage: When misbehavior is detected in the checking stage, the processors in
$P_{match}$ are required to {\em broadcast} the coded symbol they sent in the matching stage, using the
\BSB algorithm. Using the information received during these broadcasts, the fault-free processors are able to
learn new information regarding the potential identity of the faulty processor(s).
The {\em diagnosis graph} (called \DG in Algorithm \ref{alg:consensus}) is updated to incorporate
this new information.
\end{enumerate}

In the rest of this section, we discuss each of the three stages in more detail.
Note that whenever algorithm \BSB is used, all the fault-free processors will receive
the broadcasted information identically. One instance of \BSB is needed for
each bit of information broadcasted using \BSB.

%

\begin{algorithm}[!ht]
\caption{Multi-Valued Consensus (generation $g$)}\label{alg:consensus}
\begin{enumerate*}
\item {\bf Matching Stage: }\\
Each processor $P_i$ performs the matching stage as follows:
\begin{enumerate*}
\item Compute $(S_{i}[1],\dots,S_{i}[n]) = C_{2t}(v_i(g))$, and {\em send} $S_{i}[i]$ to every trusted processor  $P_j$
\item \label{step:send_S}
$
R_{i}[j] \leftarrow \left\{ 
\begin{array}{l}
\textrm{symbol that $P_i$ receives from $P_j$, if $P_i$ trusts $P_j$;}\\
\perp, \textrm{otherwise}
\end{array}
\right.
$

\item \label{step:M}
If $S_{i}[j] = R_{i}[j]$ then $M_{i}[j] \leftarrow $ \TRUE; else $M_{i}[j] \leftarrow $ \FALSE
\item $P_i$ broadcasts the vector $M_i$ using \BSB
\end{enumerate*}
Using the received $M$ vectors:
\begin{enumerate*}
\setcounter{enumii}{4}
\item Find a set of processors $P_{match}$ of size $n-t$ such that \\
	\hspace*{0.3in} $M_{j}[k]=M_{k}[j]=$ \TRUE for every pair of $P_j,P_k\in P_{match}$
\item If $P_{match}$ does not exist then decide on a default value and terminate;\\ else enter the Checking Stage
\end{enumerate*}

\item {\bf Checking Stage:}\\
Each processor $P_j\notin P_{match}$ performs steps 2(a) and 2(b):
\begin{enumerate*}
\item If $R_{j}/P_{match}\in C_{2t}$ then $Detected_j \leftarrow$ \FALSE; else  $Detected_j \leftarrow$ \TRUE.
\item Broadcast $Detected_j$ using \BSB.
\end{enumerate*}
Each processor $P_i$ performs step 2(c):
\begin{enumerate*}
\setcounter{enumii}{2}
\item \label{step:no_detect}
Receive $Detected_j$ from each processor $P_j\notin P_{match}$ (broadcasted in step 2(b)).\\
If $Detected_j=$ \FALSE for all $P_j\notin P_{match}$, then decide on $v_i'(g) = C^{-1}_{2t}(R_{i}/P_{match})$; \\ else enter Diagnosis Stage
\end{enumerate*}

\item {\bf Diagnosis Stage: }\\
Each processor $P_j\in P_{match}$ performs step 3(a):
\begin{enumerate*}
\item Broadcast $S_{j}[j]$ using \BSB \\ (one instance of \BSB is needed for each bit of $S_j[j]$) 
\end{enumerate*}
Each processor $P_i$ performs the following steps:
\begin{enumerate*}
\setcounter{enumii}{1}
\item $R^\#[j]\leftarrow$ symbol received from $P_j\in P_{match}$ as a result of broadcast in step 3(a)
\item For all $P_j\in P_{match}$,\\
\hspace*{0.3in} if $P_i$ trusts $P_j$ and $R_i[j]= R^\#[j]$ then  $Trust_i[j]\leftarrow $ \TRUE;\\ \hspace*{0.3in} else $Trust_i[j]\leftarrow $ \FALSE
\item Broadcast $Trust_i/P_{match}$ using \BSB
%

\item \label{step:remove_edge} For each edge $(j,k)$ in \DG, \\ \hspace*{0.3in} remove edge $(j,k)$ {\bf if}
$Trust_j[k]$ = \FALSE or $Trust_k[j]$ = \FALSE
\item \label{step:remove_false_detect}
If $R^\#/P_{match}\in C_{2t}$ then \\
\hspace*{0.3in} if for any $P_j\notin P_{match}$, \\
\hspace*{0.6in} $Detected_j =$ \TRUE, but no edge at vertex $j$ was removed in step 3(e) \\
\hspace*{0.3in} then remove all edges at vertex $j$ in \DG
\item \label{step:remove_faulty}
If at least $t+1$ edges at any vertex $j$ have been removed so far,\\ then processor $P_j$ 
must be faulty, and all edges at $j$ are removed.
\item Find a set of processors $P_{decide} \subset P_{match}$ of size $n-2t$ in the updated $Diag\_Graph$,\\ such that every pair of $P_j,P_k\in P_{decide}$ trust each other.
\item Decide on $v_i'(g) = C_{2t}^{-1}(R^\#/P_{decide})$.

\end{enumerate*}
\end{enumerate*}
\end{algorithm}

\clearpage

\subsection{Matching Stage}

The line numbers referred below correspond to the line numbers
for the pseudo-code in Algorithm \ref{alg:consensus}.

\noindent
{\tt Line 1(a):}
In generation $g$,
each processor $P_i$ first encodes $v_i(g)$, represented by $n-t$ symbols,
into a codeword $S_i$ from the code $C_{2t}$. The $j$-th symbol in the codeword is
denoted as $S_i[j]$. Then processor $P_i$ sends $S_i[i]$, the $i$-th symbol of
its codeword, to all the other processors {\em that it trusts}.
Recall that $P_i$ trusts $P_j$ if and only if there is an edge between
the corresponding vertices in the diagnosis graph (referred as \DG in the pseudo-code).

\noindent
{\tt Line 1(b):}
Let us denote by $R_i[j]$ the symbol that $P_i$ receives from
a trusted processor $P_j$. Processor $P_i$ {\bf ignores} any messages received from untrusted processors,
treating the message as a distinguished symbol $\perp$.

\noindent
{\tt Line 1(c)}:
Flag $M_{i}[j]$ is used to record whether processor $P_i$ finds processor $P_j$'s
symbol consistent with its own local value. Specifically, the pseudo-code in line 1(c)
is equivalent to the following:
\begin{itemize}
\item When $P_i$ trusts $P_j$: If $R_{i}[j] = S_{i}[j]$, then set $M_{i}[j]=$ \TRUE; else $M_{i}[j]=$ \FALSE.
\item When $P_i$ does not trust $P_j$: $M_{i}[j]=$ \FALSE.
\end{itemize}

\noindent {\tt Line 1(d):}
As we will see later, if a fault-free processor $P_i$ does not trust another
processor, then the other processor must be faulty.
Thus
entry $M_i[j]$ in vector $M_i$ is \FALSE if $P_i$ believes that processor $P_j$ is
faulty, or that the value at processor $P_j$ differs from the value at $P_i$.
Thus, entry $M_i[j]$ being \TRUE implies that, as of this time, $P_i$ believe that
$P_j$ is fault-free, and that the value at $P_j$ is possibly identical to the
value at $P_i$.
Processor $P_i$ uses \BSB to broadcast $M_i$ to all the processors. One instance of \BSB is needed
for each bit of $M_i$.

\noindent{\tt Lines 1(e) and 1(f):}
Due to the use of \BSB, all fault-free processors receive identical
vector $M_j$ from each processor $P_j$. Using these $M$ vectors,
each processor $P_i$ attempts to find a set $P_{match}$ containing exactly
$n-t$ processors such that, for every pair
$P_j,P_k\in P_{match}$, $M_{j}[k]=M_{k}[j]=$ \TRUE.
Since the $M$ vectors are received identically by all the fault-free processors
(using \BSB), they can compute identical $P_{match}$.
However, if such a set $P_{match}$ does not exist, then the fault-free
processors conclude that all the fault-free processors do not have identical
input -- in this case, they decide on some default value, and terminate the
algorithm. In the following discussion, we will show the correctness of this
step.


In the proof of the lemmas \ref{lm:matching_clique} and \ref{lm:matching_same_input},
we assume that the fault-free processors (that is, the processors in set $P_{good}$)
always trust each other -- this assumption will be shown to be correct later
in Lemma \ref{lm:diagnosis_graph}. 

\begin{lemma}
\label{lm:matching_clique}
If for each fault-free processor $P_i\in P_{good}$, $v_i(g) = v(g)$, for some value $v(g)$, then a set $P_{match}$
necessarily exists (assuming that the fault-free processors trust each other).
\end{lemma}
\begin{proof}
Since all the fault-free processors have identical input $v(g)$ in generation $g$, $S_i = C_{2t}(v(g))$ for all
$P_i\in P_{good}$. Since these processors are fault-free, and trust each other,
they send each other correct messages
in the matching stage. Thus, $R_{i}[j] = S_{j}[j] = S_{i}[j]$ for all $P_i,P_j\in P_{good}$.
This fact implies that 
$M_{i}[j]=$ \TRUE for all $P_i,P_j\in P_{good}$. Since there are at least $n-t$ fault-free
processors, it follows that a set $P_{match}$ of size $n-t$ must exist.
\end{proof}

Observe that, although the above proof shows that there exists a set $P_{match}$ containing
only fault-free processors, there may also be other such sets that contain some faulty processors
as well. That is, all the processors in $P_{match}$ cannot be assumed to be fault-free.

Converse of Lemma \ref{lm:matching_clique} implies that, if a set $P_{match}$ does not exist, it is certain that the fault-free processors do not have the same input values. In this case, they can correctly agree on some default value
and terminate the algorithm. This proves the correctness of Line 1(f).

In the case when a set $P_{match}$ is found, the following lemma is useful.
\begin{lemma}
\label{lm:matching_same_input}
The fault-free processors in $P_{match}$ (that is, all the processors in $P_{match}\cap P_{good}$)
have the same input for generation $g$.
\end{lemma}
\begin{proof}
$|P_{match}\cap P_{good}|\ge n-2t$ because $|P_{match}| = n-t$ and there are at most $t$ faulty processors. 
Consider any two processors $P_i,P_j\in P_{match}\cap P_{good}$. Since $M_i[j]=M_j[i]= \TRUE$,
it follows that $S_{i}[i] = S_{j}[i]$ and $S_{j}[j] = S_{i}[j]$. Since there are $n-2t$ fault-free
processors in $P_{match}\cap P_{good}$, this implies that the codewords computed by these
fault-free processors
(in Line 1(a)) contain at least $n-2t$ identical symbols. Since the code $C_{2t}$ has dimension $(n-2t)$,
this implies that the fault-free processors in $P_{match}\cap P_{good}$
must have identical input in generation $g$.
\end{proof}

\subsection{Checking Stage}

When $P_{match}$ is found during the matching stage, the checking stage is entered.

\noindent{\tt Lines 2(a) and 2(b):}
Every fault-free processor $P_j\notin P_{match}$ checks whether the symbols received from the
trusted processors in $P_{match}$ are consistent with a valid codeword: that is, check
whether $R_{j}/P_{match} \in C_{2t}$. The result of this test is broadcasted as a 1-bit notification
$Detected_i$, using \BSB.
If $R_{j}/P_{match} \notin C_{2t}$, then processor $P_j$ is said to have detected an {\em inconsistency}.

\noindent{\tt Line 2(c):}
If no processor announces in Line 2(b) that it has detected an inconsistency, each
fault-free processor $P_i$ chooses $C_{2t}^{-1}(R_i/P_{match})$ as its decision value
for generation $g$.

The following lemma argues correctness of the decision made in Line 2(c).
\begin{lemma}
\label{lm:checking}
If no processor detects inconsistency in Line 2(a), all fault-free processors $P_i\in P_{good}$ 
decide on the identical output value $v'(g)$ such that $v'(g) = v_j(g)$ for all $P_j\in P_{match}\cap P_{good}$.
\end{lemma}
\begin{proof}
Observe that size of set $P_{match}\cap P_{good}$ is at least $n-2t$,
and hence the inverse operations $C_{2t}^{-1}(R_i/P_{match})$
and $C_{2t}^{-1}(R_i/P_{match}\cap P_{good})$ are both defined.

Since fault-free processors send correct messages,
$R_{i}/P_{match}\cap P_{good}$ are identical for all fault-free processors $P_i\in P_{good}$.
Since no inconsistency has been detected by any processor, every fault-free processor $P_i$ decides on  $C_{2t}^{-1}(R_i/P_{match})$ as its output. Since $C_{2t}$ has dimension $(n-2t)$, $C_{2t}^{-1}(R_i/P_{match}) = C_{2t}^{-1}(R_i/P_{match}\cap P_{good})$.
 It then follows that all the fault-free processors $P_i$  decide
on the identical value $v'(g) =  C_{2t}^{-1}(R_i/P_{match}\cap P_{good})$
in Line 2(c). Since $R_j/P_{match}\cap P_{good} = S_j/P_{match}\cap P_{good}$ for all processors $P_j\in P_{match}\cap P_{good}$, $v'(g) = v_j(g)$ for all $P_j\in P_{match}\cap P_{good}$.
\end{proof}

\subsection{Diagnosis Stage}
When any processor  that is not in $P_{match}$ announces that it has
detected an inconsistency, the diagnosis stage is entered.
The algorithm allows for the possibility that
a faulty processor may erroneously announce that it has detected an
inconsistency.
The purpose of the diagnosis stage is to learn new information regarding
the potential identity of a faulty processor. The new information is
used to remove one or more edges from the diagnosis graph \DG -- as we will soon
show, when an edge $(j,k)$ is removed from the diagnosis graph, at least
one of $P_j$ and $P_k$ must be faulty. We now describe the steps in the
Diagnosis Stage.

\noindent{\tt Lines 3(a) and 3(b):} Every fault-free processor $P_j\in P_{match}$ uses \BSB to broadcast $S_{j}[j]$ to all processors. Let us denote by $R^\#[j]$ the result of the broadcast from $P_j$. Due to the use of \BSB, all fault-free processors receive identical $R^\#[j]$ for each processor $P_j\in P_{match}$.
This information will be used for diagnostic purposes.
%
%

\noindent{\tt Line 3(c) and 3(d):} Every fault-free processor $P_i$ uses  flag $Trust_i[j]$ to record whether  it ``{\em believes}'', as of this time, that each processor  $P_j\in P_{match}$ is fault-free or not. Then $P_i$ broadcasts $Trust_i/P_{match}$ to all processors using  \BSB. Specifically, 
\begin{itemize}
\item If $P_i$ trusts $P_j$ {\bf and} $R_i[j] = R^\#[j]$, then set $Trust_i[j]=$ \TRUE; 
\item If $P_i$ does not trust $P_j$ {\bf or} $R_i[j] \neq R^\#[j]$, then set $Trust_i[j]=$\FALSE.
\end{itemize}

\noindent{\tt Line 3(e):} Using the $Trust$ vectors, each fault-free processor $P_i$ then removes any edge $(j,k)$ from the diagnosis graph such that $Trust_j[k]$ or $Trust_k[j]=$ \FALSE. Due to the used of \BSB, all fault-free processors receive identical  $Trust$ vectors. Hence they will remove the same set of edges and maintain an identical view of the updated \DG. 

\noindent{\tt Line 3(f):} As we will soon show, in the case $R^\#/P_{match}\in C_{2t}$, a processor $P_j\notin P_{match}$ that announces that it has detected an inconsistency, i.e., $Detected_j =$\TRUE, must be faulty if no edge attached to vertex $j$ was removed in Line 3(e). Such processors $P_j$ are ``{\em isolated}'', by having  all edges attached to vertex $j$ removed from \DG, and the fault-free processors will not communicate with it anymore in subsequent generations.


\noindent{\tt Line 3(g):} As we will soon show, a processor $P_j$ must be faulty if at least $t+1$ edges at vertex $j$ have been removed.  The identified faulty processor $P_j$ is then isolated.

\noindent{\tt Lines 3(h) and 3(i):} Since \DG is updated only with information broadcasted with \BSB ($Detected$, $R^\#$ and $Trust$), all fault-free processors maintain an identical  view of the updated \DG. Then they can compute an identical set $P_{decide}\subset P_{match}$ containing exactly $n-2t$ processors such that every pair $P_j,P_k\in P_{decide}$ trust each other. Finally, every fault-free processor chooses $C_{2t}^{-1}(R^\#/P_{decide})$ as its decision value for generation $g$.

We first prove the following property of the evolution of \DG.
\begin{lemma}
\label{lm:diagnosis_graph}
Every time the diagnosis stage is performed, at least one edge attached to a vertex corresponding to a faulty processor will be removed from $Diag\_Graph$, and only such edges will be removed.
\end{lemma}
\begin{proof}
 We prove this lemma by induction. For the convenience of discussion, let us say an edge $(j,k)$ is ``{\em bad}'' if at least one of $P_j$ and $P_k$ is faulty. 

Consider  a generation $g$ starting with any instance of the \DG in which only bad edges have been removed.
When the diagnosis stage is performed, there are two possibilities: (1) a fault-free processor $P_i\notin P_{match}$ detects an inconsistency; or (2) a faulty processor  $P_j\notin P_{match}$ announces that it has detected an inconsistency. We consider the two possibilities separately:
\begin{enumerate}
\item A fault-free processor  $P_i\notin P_{match}$ detects an inconsistency: In this case, $R_i/P_{match}\notin C_{2t}$. However, according to the definition of $P_{match}$, $R_k/P_{match}=S_k/P_{match}\in C_{2t}$ for every processor $P_k\in P_{match}\cap P_{good}$. This implies that there must be a faulty processor $P_j\in P_{match}$, which is trusted by $P_i$ and $P_k$, has sent different symbols to the fault-free processors $P_i$ and $P_k$ during the matching stage. Thus, the $R^\#[j]$ must be different from at least one of   $R_i[j]$ and $R_k[j]$. As a result, $Trust_i[j]=$ \FALSE or $Trust_k[j]=$ \FALSE. Then at least one of the bad edges $(i,j)$ and $(j,k)$  will be removed in Line 3(e).

\item A faulty processor $P_j\notin P_{match}$ announces that it detects an inconsistency: Denote by $X\subset P_{match}$ the set of processors $\in P_{match}$ that $P_j$ trusts. According to the algorithm, either an bad edge $(j,k)$ for some $P_k\in X$ was removed in Line 3(e), or none of such edges is removed. In the former case, the bad edge $(j,k)$ is removed. In the later case, there are two possibilities
\begin{enumerate}
\item $R^\#/{P_{match}}\in C_{2t}$: Given that no edge $(j,k)$ for every $P_k\in X$ was removed in Line 3(e), one can conclude that, {\em if $P_j$ is fault-free}, then $Trust_j[k]=$\TRUE for all $P_k\in X$, and $R_j[k]/X = R^\#[k]/X\in C_{2t}$.
On the other hand,
observe that $P_j$ computes $Detected_j$ by checking whether $R_j/X\in C_{2t}$, since any message from untrusted processors in $P_{match}$ should have been ignored by $P_j$ in Line 1(b). From $Detected_j=$ \TRUE, one can conclude that, {\em if $P_j$ is fault-free},  $R_j/X\notin C_{2t}$. Now we have a contradiction if $P_j$ is fault-free. So processor $P_j$ must be faulty and all edges at vertex $j$ are bad. These bad edges are removed in Line 3(f).

\item $R^\#/{P_{match}}\notin C_{2t}$: In this case, similar to the discussion in case 1,  some bad edge connecting two vertices corresponding to processors in $P_{match}$ is removed in Line 3(e).
\end{enumerate}
\end{enumerate}

So by the end of Line 3(f), at least one new bad edge has been removed. Moreover, since $R_i[k] = R^\#[k]$ for all fault-free processors $P_k \in P_{match}\cap P_{good}$, $Trust_{i}[k]$ remains \TRUE for every pair of processors $P_i,P_k\in P_{good}$, which implies that the vertices corresponding to the fault-free processors will remain fully connected, and each will always have at least $n-t-1$ edges. This follows that a processor $P_j$ must be faulty if at least $t+1$ edges  at vertex $j$ has been removed. So all edges at $j$ are bad and will be removed in Line 3(g).

Now we have proved that for every generation that begins with a $Diag\_Graph$ in which only bad edges have been removed, at least one new bad edge, and only bad edges, will be removed in the updated $Diag\_Graph$ by the end of the diagnosis stage. Together with the fact that $Diag\_Graph$ is initialized as a complete graph, we finish the proof.
\end{proof}

The above proof of Lemma \ref{lm:diagnosis_graph} shows that all fault-free processors will trust each other throughout the execution of the algorithm, which justifies the assumption made in the proofs of the previous lemmas. The following lemma shows the correctness of Lines 3(h) and 3(i).
\begin{lemma}
\label{lm:diagnosis_consensus}
By the end of diagnosis stage, all fault-free processors $P_i\in P_{good}$ decide on the same output value $v'(g)$, such that $v'(g) = v_j(g)$ for all $P_j\in P_{match}\cap P_{good}$.
\end{lemma}
\begin{proof}
First of all, the set $P_{decide}$ necessarily exists since there are at least $n-2t\ge t+1$ fault-free processors in $P_{match}\cap P_{good}$ that always trust each other. Secondly, since the size of  $P_{decide}$ is $n-2t\ge t+1$, it must contain at least one fault-free processor $P_k\in P_{decide}\cap P_{good}$. Since $P_k$ still trusts all processors of $P_{decide}$ in the updated $Diag\_Graph$, $R^\#/{P_{decide}} = R_k/P_{decide}= S_k/P_{decide}$. The second equality is due to the fact that $P_k\in P_{match}$. Finally, since the size of set $P_{decide}$ is $n-2t$, the inverse operation of 
 $C^{-1}_{2t}	(R^\#/{P_{decide}})$ is defined, and it equals to
 $C^{-1}_{2t}	(S_k/{P_{decide}}) = v_k(g) = v_j(g)$ for all $P_j\in P_{match}\cap P_{good}$, as per Lemma \ref{lm:matching_same_input}.
\end{proof}

We can now conclude the correctness of the Algorithm \ref{alg:consensus}.
\begin{theorem}
\label{thm:consensus}
Given $n$ processors with at most $t<n/3$ are faulty, each given an input value of $L$ bits, Algorithm \ref{alg:consensus} achieves consensus correctly in  $ L/D$ generations
, with the diagnosis stage performed for at most $t(t+1)$ times.
\end{theorem}
\begin{proof}
According to Lemmas \ref{lm:matching_clique} to \ref{lm:diagnosis_consensus}, consensus is achieved correctly for each generation $g$ of $D$ bits. So the termination and consistency properties are satisfied for the $L$-bit outputs after $L/D$ generations. Moreover, in the case all fault-free processors are given an identical $L$-bit input $v$, the $D$ bits output $v'(g)$ in each generation $g$ equals to $v(g)$ as per Lemmas \ref{lm:matching_clique}, \ref{lm:checking} and \ref{lm:diagnosis_consensus}. So the $L$-bit output $v'=v$ and the validity property is also satisfied.

According to Lemma \ref{lm:diagnosis_graph} and the fact that a faulty processor $P_j$ will be removed once more than $t$ edges at vertex $j$ have been  removed, it takes at most $t(t+1)$ instance of the diagnosis stage before all faulty processors are identified. After that, the fault-free processors will not communicate with the faulty processors. Thus, the diagnosis stage will not be performed any more. So it will be performed for at most $t(t+1)$ times in all cases.
\end{proof}

\subsection{Complexity}
We have  discussed the operations of the proposed multi-valued consensus algorithm above. Now let us study the communication complexity of this algorithm. Let us denote by $B$ the complexity of broadcasting 1 bit with one instance of \BSB. In every generation, the complexity of each stage is as follows:
\vspace{-5pt}
\begin{itemize}
\item Matching stage: every processor $P_i$ sends at most $n-1$ symbols, each of $D/(n-2t)$ bits, to the processors that it trusts, and broadcasts $n-1$ bits for $M_i$. So at most 
$\frac{n(n-1)}{n-2t}D + n(n-1)B$ bits
in total are  transmitted by all $n$ processors.


\item Checking stage: every processor $P_j\notin P_{match}$ broadcasts one bit $Detected_j$ with \BSB, and there are $t$ such processors. So $tB$ bits are transmitted.

\item Diagnosis stage: every processor $P_j \in P_{match}$ broadcasts one symbol $S_j[j]$ of $D/(n-2t)$ bits with \BSB; and every processor $P_i$ broadcasts $n-t$ bits of $Trust_i/P_{match}$ with \BSB. So the complexity  is
$\frac{n-t}{n-2t}DB + n(n-t)B$ bits.
\end{itemize}

According to Theorem \ref{thm:consensus}, there are $L/D$ generations in total. In the worst case, $P_{match}$ can be found in every generation, so the matching and checking stages will be performed for $L/D$ times. In addition, the diagnosis stage will be performed for at most $t(t+1)$ time. Hence  the communication complexity of the proposed consensus algorithm, denoted as $C_{con}(L)$, is then computed as
\begin{eqnarray}
C_{con}(L) &=&\left(\frac{n(n-1)}{n-2t}D + n(n-1)B + tB\right)\frac{L}{D}  + t(t+1)\left(\frac{n-t}{n-2t}D + n(n-t)\right)B
\end{eqnarray}
%
For a large enough value of $L$, with a suitable choice of 
$D = \sqrt{\frac{(n^2-n+t)(n-2t)L}{t(t+1)(n-t)}}$,
we have
\begin{eqnarray}
C_{con}(L) &=& \frac{n(n-1)}{n-2t}L  + 2BL^{0.5}\sqrt{\frac{(n^2-n+t)t(t+1)(n-t)}{n-2t}} + t(t+1)n(n-t)B
\end{eqnarray}

Error-free algorithms that broadcast 1 bit with communication complexity $\Theta(n^2)$ bits are known \cite{bit_optimal_89,opt_bit_Welch92}. So we assume $B=\Theta(n^2)$. Then the complexity of our algorithm for  $t<n/3$ becomes
\begin{equation}
C_{con}(L) = \frac{n(n-1)}{n-2t}L + O(n^4L^{0.5}  + n^6) = O(nL + n^4 L^{0.5}  + n^6).
\end{equation}
So for sufficiently large $L$ ($\Omega(n^6)$), the communication complexity approaches $O(nL)$.

%% file: broadcast_short.tex
%
%

\section{Multi-Valued Broadcast and Tolerating $t\ge n/3$ Failures}
\label{sec:broadcast}
Here we briefly discuss the Byzantine {\em broadcast} problem (also known as the ``Byzantine Generals Problem'' \cite{psl82}). Similar to the consensus problem, the broadcast problem also considers achieving agreement among $n$ processors: A designated {\em ``source''} processor tries to broadcast an $L$-bit value to the other processors, while $t<n/3$ processors (probably including the source) may be faulty. Using techniques introduced in this paper, we can achieve error-free multi-valued broadcast with communication complexity 
$C_{bro}(L) <1.5(n-1)L + \Theta(n^4L^{0.5})$ bits
for $t<n/3$ and large $L$ \cite{techreport_BA_complexity}. Notice that the complexity of any broadcast algorithm, even the ones that allow a positive probability of error, is lower bounded by $(n-1)L$. So we can achieve error-free broadcast with complexity within a factor of $1.5+\epsilon$ to the optimal for any constant $\epsilon>0$ and sufficiently large $L$.

%% file: conclusion.tex
Most of our discussion in the previous section is independent of the number of faulty processors. The requirement for $t<n/3$ is needed only for the correctness of the deterministic error-free 1-bit broadcast algorithm \BSB. In practice, it may be desirable to be able to tolerate  $t\ge n/3$ failures at the cost of a non-zero probability of error. This need can be met by our algorithm with a small modification: substitute \BSB with any probabilistically correct 1-bit broadcast algorithm that tolerates the desired number of failures (ones with authentication from \cite{Waidner96information-theoreticpseudosignatures,authenticated_BA_Dolev83} for example). With this modification, our algorithm tolerates the same number of failures as the 1-bit broadcast algorithm does, and makes an error only if the 1-bit broadcast algorithm fails. The only difference in the communication complexity is the term {\em sub-linear} in $L$. So for sufficiently large $L$, the complexity of the modified algorithm is also $O(nL)$.

\section{Conclusion}
\label{sec:conclusion}

In this paper, we present efficient error-free Byzantine consensus algorithm for long messages. The algorithm requires $O(nL)$ total bits of communication for messages of $L$ bits for sufficiently large $L$. Our algorithm makes no cryptographic assumption and still is able to always solve the Byzantine consensus problem correctly. 